\documentclass[aps,prl,twocolumn,showpacs,amsmath,amssytmb]{revtex4}
\usepackage{graphicx}
\usepackage{dcolumn}
\usepackage{bm}
\usepackage{multirow}

\begin{document}


\title{
New Identification of Metallic Phases of In Atomic layers on Si(111) Surfaces
}

\author{Kazuyuki Uchida}

\author{Atsushi Oshiyama}

\affiliation{Department of Applied Physics, The University of Tokyo, 
Tokyo 113-8656, Japan}


\date{\today}

\begin{abstract}
We report first-principles calculations that clarify atomic structures and coverage of the metallic phases of In overlayers on Si (111) surfaces. Calculated energy bands and scanning tunneling microscopy images along with the obtained energetics of various phases reveal that the two metallic phases with the $\sqrt{7} \times \sqrt{3}$ periodicity observed experimentally are single and double In overlayers, as opposed to prevailing assignments. 
\end{abstract}


                              
\maketitle


Surfaces ubiquitous in nature provide new phases of materials due to symmetry breaking and modification of interactions among constituting elements. A thin layer of metallic elements on a semiconductor surface is an example. It offers a good stage to study two-dimensional electron systems (2DESs), and also a key structure in device fabrication in current technology encountered with the cutting-edge miniaturization \cite{itrs}. 

Indium 
adatom layers on Si(111) surfaces are typical and important examples. A particular phase with the lateral periodicity of $\sqrt{7} \times \sqrt{3}$ exhibits metallic behavior down to several K \cite{hasegawa}, as opposed to the metal-insulator transition predicted for the 2DESs \cite{Anderson, MI}, and eventually becomes superconducting
at 3 K \cite{zhang, uchihashi}. 
This is the first case where 
superconductivity is found in a deposited layer.
Angle-resolved photoemission spectroscopy (ARPES) clarifies the  
band
structure of the metallic phase of the 
$\sqrt{7} \times \sqrt{3}$ surface \cite{rotenberg}. 
However, atom-scale identification, {\it i.e.}, the In coverage, the stable atomic structure and the resultant electron states, of the metallic phase is still lacking. 

Various surface reconstructions emerge 
by depositing In atoms 
on the Si(111) 7 $\times$ 7 surface \cite{review}, 
followed by annealing at $\sim$ 500 $^\circ$C
in ultra-high vacuum: 
The $\sqrt{3} \times \sqrt{3}$, 
the $\sqrt{31} \times \sqrt{31}$, 
and the $4 \times 1$ phases appear 
consecutively with increasing the In dose, 
all showing insulating behavior 
at low temperature \cite{kraft1, yeom}. 
Then the $\sqrt{7} \times \sqrt{3}$ phase \cite{kraft0} appears 
with the In deposition of 1.5 - 1.8 ML. 
The STM topographs \cite{kraft1, kraft0} 
show that the two distinctive structures coexist 
on the $\sqrt{7} \times \sqrt{3}$ phase: 
Bright spots appear in a 
quasi-hexagonal pattern with protruding trimers 
in one structure (hex structure hereafter), 
and they appear in a quasi-rectangular pattern 
in the other structure (rect structure). 
The former and the latter have been speculated 
to be 1.0 ML and 1.2 ML, respectively, 
In adatoms on the Si(111) surface \cite{kraft1, kraft0}. 
The excess In atoms are
desorbed from the surface
in the annealing.
The metallic behavior \cite{hasegawa, rotenberg} 
and the superconducting gap \cite{zhang} 
have been observed on the rect structure, 
whereas the metallic behavior and the superconducting current has been measured on the hex structure \cite{uchihashi}. Identification of 
the hex and the rect structures 
based on the first-principles calculations is 
highly demanded. 

In this Letter, 
we unequivocally identify 
the two metallic phases of $\sqrt{7} \times \sqrt{3}$ - In/Si(111) 
by performing total-energy
electronic-structure calculations in the density-functional theory. 
The obtained energetics,
energy bands and scanning tunneling microscopy (STM) images 
are indicative that the surface measured by ARPES 
is of the rect structure and 
is the 2.4 monolayer (ML) In atomic layers, in sharp contrast to the possibilities discussed 
in the past. 
The hex structure, on the other hand, is clearly identified as the 1.2 ML In atomic layers. We have found that the both structures are metallic. 

Our calculations have been performed with using the local density approximation (LDA) \cite{lda} 
in 
the density functional theory (DFT) \cite{hohenberg, kohn}. Norm-conserving pseudopotentials 
\cite{troullier,kleinman} 
are used to describe the electron-ion interactions. 
Valence-state Kohn-Sham (KS) wave functions 
are expanded by a plane-wave basis set 
with the cut-off energy of 49 Ryd. 
The In-covered Si(111) surface is simulated by a 
repeated 
slab model. 
Geometry optimization is done 
until the remaining force on each atom becomes less than 50 meV / \AA \cite{details}. 

\begin{figure}
\includegraphics[width=1.00\linewidth]{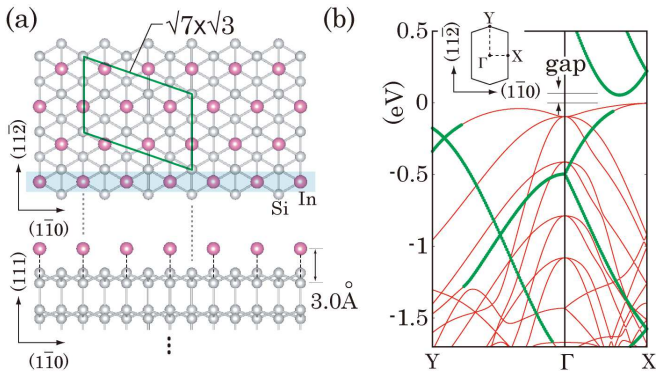}
\caption{(color online) 
(a) Top and side views of the stable structure of the 1.0 ML In surface. 
Pink (large) and silver (small) balls depict 
In and Si atoms, respectively. Only the In atoms in the shaded region of the top view are 
shown 
in the side view. 
(b) Band structure with the lateral Brillouin zone corresponding to the $\sqrt{7} \times \sqrt{3} $ lateral cell. 
The Kohn-Sham (KS) electron states which have the character of In orbitals more than 50 \% are depicted by 
green (bold) lines. 
}
\label{cov1.0}
\end{figure}

We start with the 1.0 ML In on Si(111) in which there are 5 In atoms per $\sqrt{7} \times \sqrt{3}$ lateral cell. This has been postulated to be the hex structure observed by STM. 
We prepare several initial geometries where In atoms are located at 
plausible 
positions in the cell, and perform geometry optimizations. 
We have then reached the most stable 
structure in which the $\sqrt{7} \times \sqrt{3}$ periodicity disappears and the 1 $\times$ 1 periodicity comes up instead [Fig. \ref{cov1.0}(a)]. 
The next stable structure 
keeps the $\sqrt{7} \times \sqrt{3}$ periodicity, 
but its total energy is higher than the most stable structure
by 50 meV per 1 $\times$ 1 area. 
Figure \ref{cov1.0}(b) shows the energy bands of the most stable structure. Hybridization of In orbitals with Si dangling-bond orbitals 
leads to a {\it non}-metallic band structure with the energy gap of $\sim 50$ meV. 
The hex surface identified by the STM experiment is known 
to be metallic, however
\cite{uchihashi}.
We thus 
conclude that the In 1.0 ML surface is {\it not} the observed $\sqrt{7} \times \sqrt{3}$-hex structure. 

\begin{figure}
\includegraphics[width=1.00\linewidth]{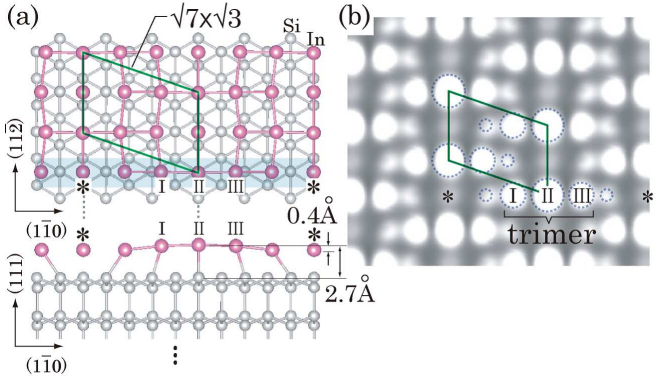}
\caption{(color online) 
Top and side views
of the stable structure for
the 1.2 ML In surface.
Pink (large) and silver (small) 
balls 
show In and Si atoms, respectively. 
(b) Calculated STM image.
Dotted circles 
are guides for eye, showing bright spots. 
Trimers come from upper In atoms labeled as I, II, and III.
The In atoms labeled as $\ast$ 
become 
invisible in the STM image.
}
\label{cov1.2}
\end{figure}
\begin{figure}
\includegraphics[width=1.00\linewidth]{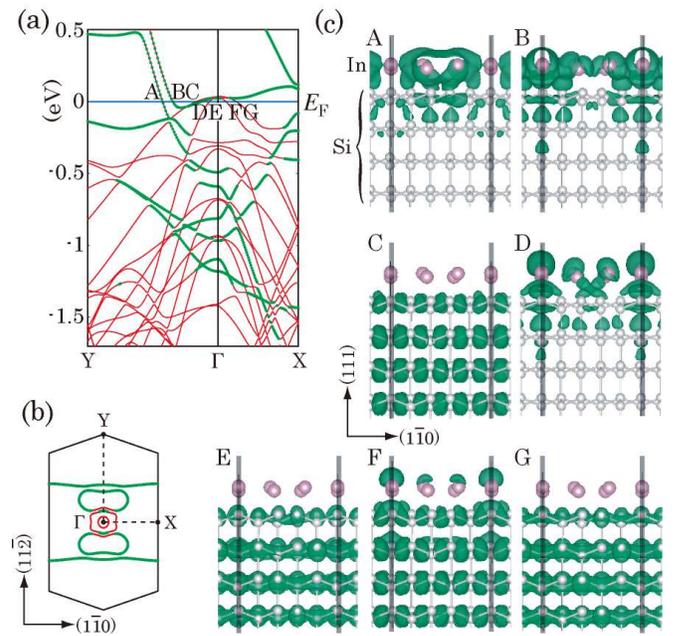}
\caption{(color online) 
Energy bands (a) 
and Fermi lines (b)
of the 1.2 ML In surface.
The
KS 
electron states which have the character of In orbitals more than 50 \% are depicted by 
green (bold) lines. 
The Fermi level $E_{\rm F}$ is set to be 0. 
(c) Isosurface of the squared absolute values of 
the
KS 
orbitals
on the Fermi level,
at 10 \% of the maximum.
}
\label{cov1.2bands}
\end{figure}

We next examine the 1.2 ML In on the Si(111), 
which has been postulated to be the rect structure observed by STM. The number of In atoms is 6 per cell. 
In the obtained 
stable
geometry, 
the 3 In atoms
are located above the top-layer Si
atoms, 
whereas others lack the partner Si atoms, 
leading to a corrugation of the In layer as large as 0.4 {\AA} [Fig. \ref{cov1.2}(a)]. 
The corrugartion is indeed observed by STM for both the hex and the rect structures \cite{kraft0}. 

Figure \ref{cov1.2}(b) 
shows
the calculated STM image of the 1.2 ML In on Si(111) \cite{stm}. 
We clearly observe bright trimers which reflect the electron density from the upper 3 In atoms 
in the corrugated structure. The corrugation also makes 
certain 
In atoms, labeled as 
$\ast$ 
in Fig. \ref{cov1.2}(a) 
sink to the Si surface and invisible in the STM image. 
Then the calculated STM image shows 5, {\it not} 6, bright spots with 
a 
prominent trimer in the $\sqrt{7} \times \sqrt{3}$ cell. This feature is exactly what is observed experimentally for the hex structure. 
We thus naturally conclude that the hex structure is the 1.2 ML In on Si(111), as opposed to the previous assignment to the 1.0 ML In on Si(111). 

Figure \ref{cov1.2bands} 
shows calculated energy bands (a) 
and 
Fermi lines (b) 
of the 1.2 ML In surface. 
We find that the band structure has a 
metallic 
character with several bands crossing 
the Fermi level $E_{\rm F}$. 
Two of them, shown by red (thin) lines in Fig. \ref{cov1.2bands}(a),  
have a character of the 
top of the Si valence 
bands, and
the others,
shown by green (bold) lines in Fig. \ref{cov1.2bands}(a),   
are of In characters.
The corresponding wave functions are
shown in Fig. \ref{cov1.2bands}(c). 
The calculated energy bands are 
qualitatively different from 
those determined by ARPES 
experiment for the rect structure (Fig. 2(c) in Ref. \cite{rotenberg}). 
This clearly indicates that the rect structure which has been investigated by the ARPES and is also assumed to be responsible for the electron transport at low temperature \cite{hasegawa,zhang} is {\em not} the 1.2 ML In on Si(111).

\begin{figure}
\includegraphics[width=1.00\linewidth]{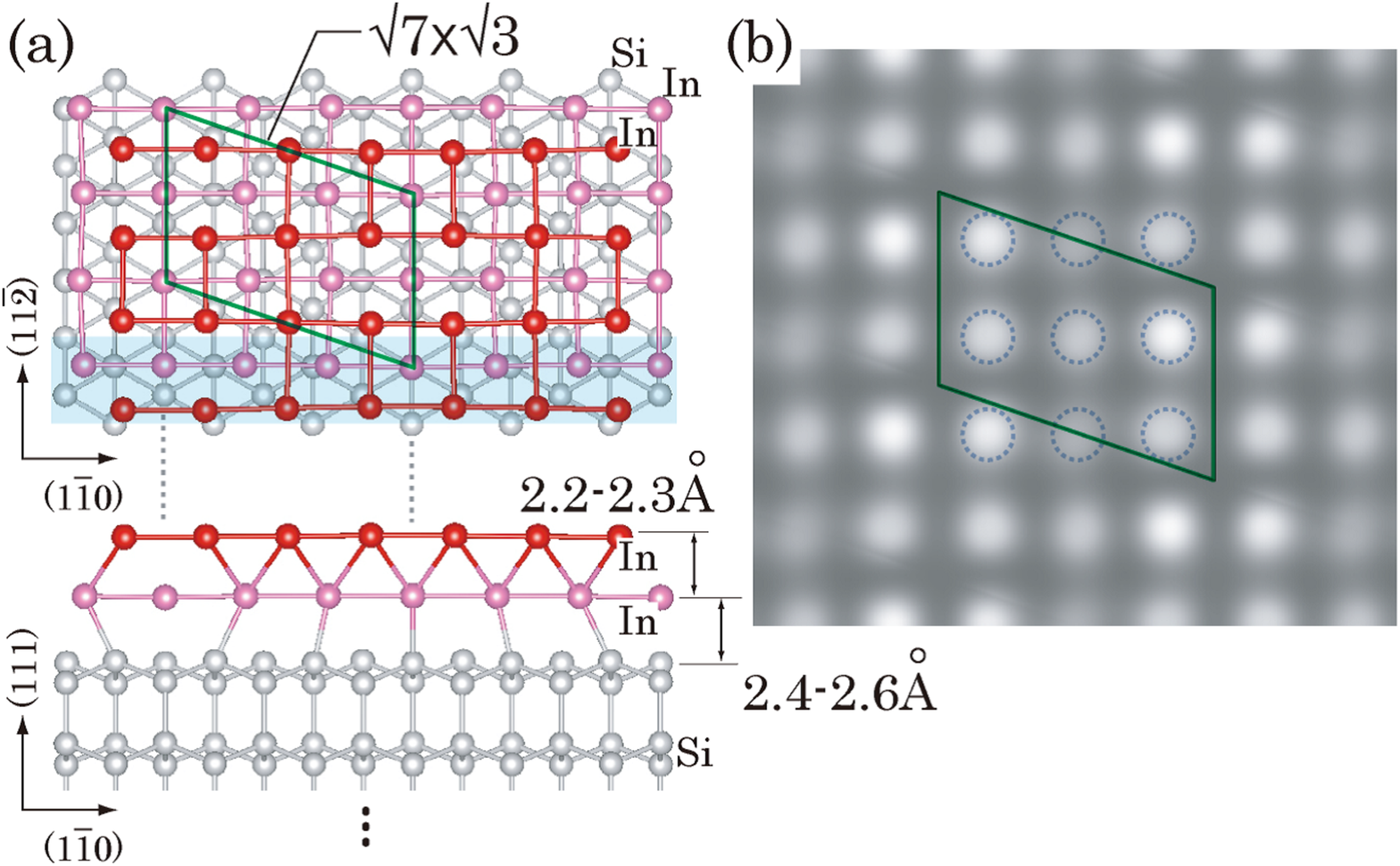}
\caption{(color online) 
(a) 
Top and side views of 
the stable structure 
of 2.4 ML In surface.
Red and pink balls are In atoms in the top and
the 2nd layers, respectively, and silvers are Si atoms. 
(b) 
Calculated STM image.
Dotted circles are guides for eyes showing bright spots. 
}
\label{cov2.4}
\end{figure}

Then, what is the identity of the rect structure? 
We now argue that it is a double layer In atoms on Si(111) 
with the coverage of 2.4 ML. 
Figure \ref{cov2.4}(a) shows the most stable 
structure of the 2.4 ML In on Si(111).
The number of In atoms per $\sqrt{7} \times \sqrt{3}$ cell is 12, 
of which 6 are in the top In layer and the other 6 are in the second In layer. 
This arrangement is akin to that on the (001) face of 
the body centered tetragonal In crystal.  
The calculated amount of the corrugation in the top In layer is 0.1 {\AA}, which is substantially smaller than the value calculated 
for the stable 1.2 ML In structure (0.4 {\AA}).
It is of note that, in the STM measurements, the corrugation in the rect structure is 
smaller than that in the hex structure.

\begin{figure}
\includegraphics[width=1.00\linewidth]{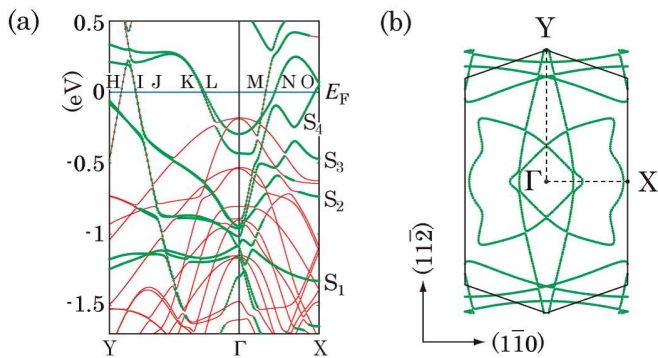}
\caption{(color online)
Energy bands (a) and Fermi lines (b) 
of the 
2.4 ML In surface.
The KS electron states 
which have the character of In orbitals more than 50 \% are depicted by 
green (bold) lines.
}
\label{cov2.4bands}
\end{figure}

\begin{figure}
\includegraphics[width=1.00\linewidth]{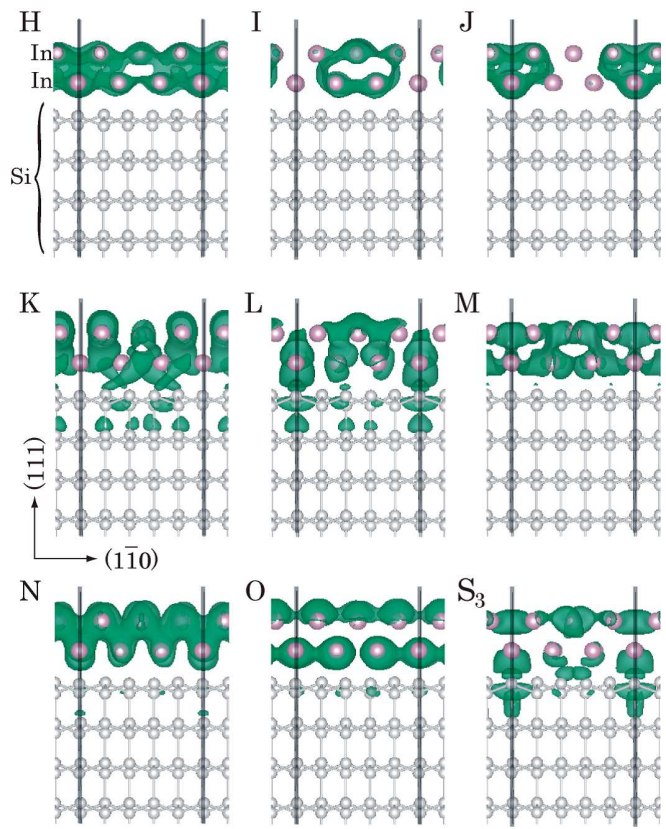}
\caption{(color online)
Isosurface of the squared absolute values 
of the KS 
orbitals 
on the Fermi level
$E_{\rm F}$
of the 
2.4 ML In surface,
at 15 \% of maximum. 
}
\label{cov2.4bands2}
\end{figure}

Figure \ref{cov2.4}(b) is the STM image 
calculated for the most stable structure of the 2.4 ML In on Si(111). We observe a clear rectangular arrangement of bright spots, which correspond to the 
positions 
of the top-layer In atoms. 
We find that the calculated STM image is 
essentially identical to the experimental one for the rect structure, 
which has been incorrectly assigned to the 1.2 ML In on Si(111) in the past. 

Figure \ref{cov2.4bands} 
shows calculated energy bands (a)
and Fermi lines (b) 
for the stable 2.4 ML In surface. 
Several bands cross the Fermi level 
$E_{\rm F}$, showing a metallic nature. 
Interestingly, the states at $E_{\rm F}$ dominantly consist of In orbitals (Fig. \ref{cov2.4bands2}), 
inferring that two-dimensional In metal is formed on Si(111). 
In Fig. \ref{cov2.4bands}(b),
we find several distinctive Fermi lines: 
the first line is located near $\Gamma$ point forming a square shape around $\Gamma$; 
the second line extends to $\Gamma$-Y direction forming a bow shape; 
the third line is around Y point extending to $\Gamma$-X direction;
the forth line exists around X point extending to $\Gamma$-Y direction 
and forming a butterfly shape.

The presence of the first, the second and the third lines 
may be understood qualitatively 
in terms of a two-dimensional 
nearly-free-electron model \cite{rotenberg}. 
However, the presence of the forth line, 
or equivalently the presence of S$_4$ band in Fig. \ref{cov2.4bands}(a), 
is unable to be explained by the simple model. 
Even for the first, the second and the third lines, 
the interaction between In and Si orbitals 
modifies the dispersion of the energy bands substantially.
Rotenberg \cite{rotenberg} reported ARPES data 
for the rect structure. 
The energy bands determined by ARPES show 
the presence of the nearly-free-electron bands 
and also other bands which are 
unable to be explained in the simple model 
(S$_1$, S$_2$ and S$_3$ in Fig. 2(c) of Ref. \cite{rotenberg}). 
Our calculations have reproduced these unidentified energy bands, 
and clarified that they are In states mixed with Si dangling-bond states
[see S$_3$ in Fig. \ref{cov2.4bands2}]. 
Agreement of our energy bands and Fermi lines with the experimental ones is
excellent \cite{ARPES}. From these electronic structures along with the STM images explained above,
we 
conclude
that the observed rect structure 
is the double-layer In on Si(111) with the coverage of 2.4 ML. 

\begin{figure}
\includegraphics[width=0.70\linewidth]{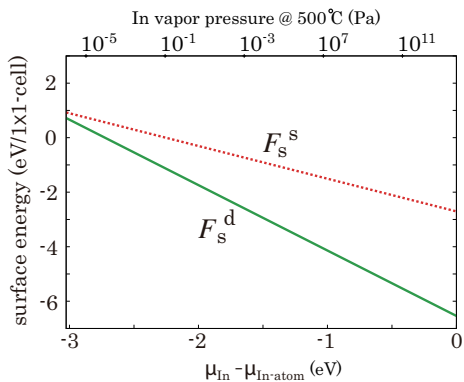}
\caption{(color online) 
Surface energy of 
a single-layer (1.2ML) In on Si(111), $F_{\rm s}^{\rm s}$, and 
that of a double-layer (2.4ML) In on Si(111), $F_{\rm s}^{\rm d}$,
as a function of the In chemical potential $\mu_{\rm In}$.
In vapor pressure at 500 $^{\circ}{\rm C}$
corresponding to the value of
$\mu_{\rm In}$ is also shown.
}
\label{fenergy}
\end{figure}

Our argument above is corroborated by the energetics obtained in the present calculations. Figure \ref{fenergy} shows the surface energy defined by $F_{\rm s} \equiv E_{\rm t} - F_{\rm s}^{\rm H} - \mu_{\rm In} N_{\rm In} - \mu_{\rm Si} N_{\rm Si} - \mu_{\rm H} N_{\rm H} $ as a function of the In chemical potential $\mu_{\rm In}$.  
Here $E_{\rm t}$ is the total energy of the slab, $F_{\rm s}^{\rm H}$ is the surface energy of the H-terminated surface, and $\mu$ and $N$ are the chemical potential and the number of 
atoms 
in the slab, respectively, of the corresponding elements \cite{surf_en}. We compare the surface energies of the single-layer (1.2 ML) In, $F_{\rm s}^{\rm s}$, and the double-layer (2.4 ML) In, $F_{\rm s}^{\rm d}$. The In chemical potential $\mu_{\rm In}$ is taken between the values of the bulk In 
metal and of the atomic In 
($\mu_{\rm In-atom}$), 
as the source of In to form the surfaces during the annealing in ultra-high vacuum condition is presumably various In clusters deposited on the Si surface. 
In case that the surface is in equilibrium with the In gas phase, the In chemical potential is translated to the In pressure at certain temperature, which is also shown in Fig. \ref{fenergy}. 
The result shows that for any value of $\mu_{\rm In}$ within this range, the double-layer (2.4 ML) In is 
energetically favorable. 
However, in the experiments, the amount of the deposited In is 
reportedly 
1.5 - 1.8 ML when $\sqrt{7} \times \sqrt{3}$ surfaces are made, which is less than 2.4 ML 
and more than 1.2 ML. 
Thus, it is 
expected that
obtained surface is a patchwork of the both structures.
This corresponds to the observed coexistence 
of the hex and the rect structures in the experiments \cite{kraft1,kraft0}.

We finally discuss the identity of the superconducting phase.
We have revealed that 
both 1.2 ML and 2.4 ML In-covered surfaces are stable 
with the $\sqrt{7} \times \sqrt{3}$ periodicity 
and show the metallic behaviors. 
The characters of the metallic states are 
different for the two cases, however: 
In the 1.2 ML case,
both of the Si valence top states 
and the In orbital states are on the Fermi level 
[Fig. \ref{cov1.2bands}(c)], 
whereas in the 2.4 ML case 
the Fermi level states 
are dominantly of In orbitals  
(Fig. \ref{cov2.4bands2}). 
As for the identification of the superconducting phase, 
controversial experiments are reported 
\cite{zhang, uchihashi}.
Considering that the superconducting transition temperature 
of In layers ($T_c \sim$ 3 K) is close to 
that in the In bulk ($T_c$ = 3.4 K), 
it is likely that the superconducting phase 
is the rect structure, {\it i.e.}, 
the double-layer In on Si(111) with 2.4 ML coverage, 
in which carriers are 
dominantly
of In character. 
However, it cannot be excluded that the hex structure becomes superconducting with the transition temperature being accidentally the same 
as the bulk value. 

In conclusion, 
we have calculated
the atomic and electronic structures of
the $\sqrt{7} \times \sqrt{3}$-In/Si(111) surfaces
by first-principles calculations based on the DFT.
Calculated band dispersions,
STM images, and
energetics
identify 
the $\sqrt{7} \times \sqrt{3}$-hex and
the $\sqrt{7} \times \sqrt{3}$-rect surfaces as
1.2 ML In and 2.4 ML In on Si(111), respectively.

\begin{acknowledgments}
We benefits from the discussion with Dr. T. Uchihashi
and Dr. S. Yamazaki.
This work was supported by the Grants-in-Aid for scientific research 
under the contract number 22104005 and the CMSI project, both 
conducted by MEXT, Japan. 
Computations were performed mainly at Supercomputer Center in ISSP, The University of Tokyo. 
\end{acknowledgments}


\begin{thebibliography}{}

\bibitem{itrs} 
International Technology Rodamap for Semiconductors, http://www.itrs.net/.
\bibitem{hasegawa}
S. Yamazaki, Y. Hosomura, I. Matsuda, R. Hobara, T. Eguchi, Y. Hasegawa, 
and S. Hasegawa, 
Phys. Rev. Lett. {\bf 106}, 116802 (2011).
\bibitem{Anderson}
P. W. Anderson, Phys. Rev. {\bf 109}, 1492 (1958). 
\bibitem{MI}
E. Abrahams. S. V. Kravchenko, and M. P. Sarachik, 
Rev. Mod. Phys. {\bf 73}, 251 (2001).
\bibitem{zhang}
T. Zhang, P. Cheng, W.-J. Li, Y.-J. Sun, G. Wang, X.-G. Zhu, K. He, 
L. Wang, X. Ma, X. Chen, Y. Wang, Y. Liu, H.-Q. Lin, J.-F. Jia, and 
Q.-K. Xue, 
Nature Phys. {\bf 6}, 104 (2010).
\bibitem{uchihashi} 
T. Uchihashi, P. Mishra, M. Aono, and T. Nakayama, 
Phys. Rev. Lett. {\bf 107}, 207001 (2011). 
\bibitem{rotenberg}
E. Rotenberg, H. Koh, K. Rossnagel, H. W. Yeom, J. Sch\"{a}fer, 
B. Krenzer, M. P. Rocha, and S. D. Kevan, 
Phys. Rev. Lett. {\bf 91}, 246404 (2003). 
\bibitem{review}
V. G. Lifshitz, A. A. Saranin, and A. V. Zetov, 
{\it Surface Phases on Silicon} (John Wiley \& Sons, West Sussex, 1994). 
\bibitem{kraft1}
J. Kraft, M. G. Ramsey, and F. P. Netzer, 
Phys. Rev. B {\bf 55}, 5384 (1997). 
\bibitem{yeom}
H. W. Yeom, S. Takeda, E. Rotenberg, I. Matsuda, 
K. Horikoshi, J. Sch\"{a}efer, C. M. Lee, S. D. Kevan, T. Ohta, 
T. Nagao, and S. Hasegawa, 
Phys. Rev. Lett. {\bf 82}, 4898, (1999). 
\bibitem{kraft0}
J. Kraft, S. L. Surnev, and F. P. Netzer, 
Surf. Sci, {\bf 340}, 36 (1995). 
\bibitem{lda}
J.~P.~Perdew and A.~Zunger, Phys. Rev. B {\bf 23}, 5048 (1981)
\bibitem{hohenberg}
P. Hohenberg and W. Kohn, Phys. Rev. {\bf 136}, B864 (1964). 
\bibitem{kohn}
W. Kohn and L. J. Sham, Phys. Rev. {\bf 140}, A1133 (1965). 
\bibitem{troullier} 
N. Troullier and J. L. Martins, Phys. Rev. B {\bf 43}, 1993 (1991). 
\bibitem{kleinman}
L. Kleinman and D. M. Bylander, Phys. Rev. Lett. {\bf 48}, 1425 (1982).

\bibitem{details}
The computation has been done with TAPP code: O. Sugino and A. Oshiyama, Phys. Rev. Lett. {\bf 68}, 1858 (1992); J. Yamauchi, M. Tsukada, S. Watanabe, and O. Sugino, Phys. Rev. B {\bf 54}, 5586 (1996); H. Kageshima and K. Shiraishi, Phys. Rev. B {\bf 56}, 14985 (1997). We use 4$d$, 5$s$ and 5$p$ as valence states of In pseudopotential. We obtain the lattice constants of body centered tetragonal In as $a = 3.19 {\rm \AA}$ and $ c = 4.95 {\rm \AA}$, which agree with the corresponding experimental values with the error of -2 \% and +0.02 \%, respectively. Our slab model consists of the In layer on an 8-atomic-layer Si with its bottom Si being terminated by H atoms. Each slab is separated by a 10-\AA \ thick vacuum. We use $ 6 \times 4 \times 1 $ $k$-point sampling for Brillouin zone integrations for the $\sqrt{7} \times \sqrt{3}$ lateral unit cell. 
The bottom H atoms and the atoms attached to them are fixed 
in the geometry optimization.

\bibitem{stm}
STM images are simulated by calculating local density of states above the surface [J. Tersoff and D. R. Hamann, Phys. Rev. B {\bf 31}, 805 (1985)]. The calculated images are insensitive to the bias voltage. 

\bibitem{ARPES}
Only 
near-surface 
states, 
shown by green (bold) lines in Fig. \ref{cov2.4bands}(a-b),  
are observed in the ARPES experiment \cite{rotenberg}.  

\bibitem{surf_en} 
The surface energy $F_{\rm s}^{\rm H}$ of the H-terminated Si (111) is estimated by using the 16-layer Si slab with the bottom and the top surfaces terminated by H atoms. Since the numbers of Si and H atoms in this slab are twice of the corresponding numbers in the slab used in this work, we obtain $ F_{\rm s}^{\rm H} = E_{\rm t}^{\prime} / 2 - \mu_{\rm Si} N_{\rm Si} - \mu_{\rm H} N_{\rm H}$ with $E_{\rm t}^{\prime}$ being the total energy of the 16-layer slab. 


\end{thebibliography}
\end{document}